\documentclass[final,twocolumn,5p]{elsarticle}
\pdfoutput=1
\usepackage{graphicx}        
\usepackage{multicol}        
\usepackage[bottom]{footmisc}
\usepackage{tablefootnote}
\usepackage{mathptmx}
\usepackage[scaled=.90]{helvet}
\usepackage{courier}
\usepackage{wrapfig}
\usepackage{multirow}
\usepackage{graphicx}
\usepackage{color,colortbl}
\definecolor{lightgray}{gray}{0.8}
\usepackage[shortlabels]{enumitem} 
\usepackage{amsmath}
\usepackage{url}
\usepackage{balance}
\newcommand{\bi}{\begin{itemize}[leftmargin=0.4cm]}
\newcommand{\ei}{\end{itemize}}
\newcommand{\be}{\begin{enumerate}}
\newcommand{\ee}{\end{enumerate}}

\newcommand{\fig}[1]{Figure~\ref{fig:#1}}
\newcommand{\eq}[1]{Equation~\ref{eq:#1}}
\usepackage{xcolor}

\definecolor{OliveGreen}{rgb}{0,0.6,0}
\definecolor{Gray}{rgb}{0.88,1,1}
\definecolor{Gray}{gray}{0.85}
\definecolor{Blue}{RGB}{0,29,193}
\definecolor{Yellow}{RGB}{204,204,0}
\usepackage[framed]{ntheorem}
\usepackage{framed}
\usepackage{tikz}
\usetikzlibrary{shadows}

\theoremclass{Lesson}
\theoremstyle{break}
\tikzstyle{thmbox} = [rectangle, rounded corners, draw=black,
  fill=Gray,  drop shadow={fill=black, opacity=1}]

\newshadedtheorem{lesson}{Result}

\begin{document}
\begin{frontmatter}
\title{Why is Differential Evolution Better than Grid Search for Tuning Defect Predictors?}
\author{Wei Fu\corref{cor1}}
\ead{wfu@ncsu.edu}
\author{Vivek Nair}
\ead{vivekaxl@gmail.com}
\author{Tim Menzies}
\ead{tim.menzies@gmail.com}
\cortext[cor1]{Corresponding author: Tel:19195345251(Wei)}
\address{Department of Computer Science, North Carolina State University, Raleigh, NC, USA}

\begin{abstract}
\textbf{Context:} One of the black arts of data mining is learning
the magic parameters that control the learners. In software analytics, at least for defect prediction,
several methods, like grid search and differential evolution(DE), have been
proposed to learn those parameters. They've both been proved to be able to 
improve learner performance.\\
\textbf{Objective:} Given the same amount of effort, we want to evaluate 
which parameter tuning method that can 
find better parameters in terms of performance scores and runtime.\\
\textbf{Methods:} This paper compares grid search to 
differential evolution, which is an evolutionary 
algorithm that makes extensive use
of stochastic jumps around the
search space. \\
\textbf{Results:} We find that the seemingly complete approach of
grid search does no better, and sometimes worse,
than the stochastic search. Yet, when repeated
20 times to check for conclusion validity, 
DE was over 210 times faster
(6.2 hours vs 54 days
for grid search when both tuning Random Forest over 17 test data sets with F-measure as optimzation objective).\\
\textbf{Conclusions:} These results are puzzling: why does  a quick partial search be just as effective
as a much slower, and much more, extensive search? To answer
that question, we turned to the theoretical optimization literature.
Bergstra and Bengio  conjecture that grid search is not more effective than more randomized searchers if the underlying
search space is inherently low dimensional. This is significant since
recent results show that defect prediction exhibits very low intrinsic dimensionality--
an observation that explains why a fast method like DE
may work as well as a seemingly more thorough grid search. This suggests, as a future research direction, that it might be possible to peek at data sets before doing any optimization in order to match the optimization
algorithm to the problem at hand.

\end{abstract}
 \end{frontmatter}
\vspace{1mm}

\pagenumbering{arabic} 

\section{Introduction}
Given the large amount of data now available to analysts,
many researchers in empirical software engineering are now turning to
automatic data miners to help them explore all that data \cite{Menzies:2013,me07b}. Those learners
come with many ``magic numbers'' which, if tuned, can generate better
predictors for particular data spaces. 
For example,  Fu et al.\cite{fu16} showed that, with tuning, the precision of software defect predictors learned from static code metrics, can grow by 20\% to 60\%.

Since tuning is so effective, it is tempting to enforce tuning for all data mining studies.
However, there is a problem: tuning can be  slow.
For example, in this paper, one of our tuners used 54 days of CPU time to
conduct 20 repeats of tuning  Random Forests\cite{brieman00} for one tuning goal with 17 test data sets.  Other researchers also comment that tuning may require
weeks, or more, of CPU time~\cite{Arcur11}.
One dramatic demonstration of the slowness of tuning (not for defect prediction) comes from
Wang et al.~\cite{Wang:Harman13} who needed 15 years of CPU to explore
9.3 million candidate configurations for  software clone detectors.

One way to address the cost of tuning data miners is to use cloud-based CPU farms. The advantage
of this approach is that it's simple to implement (just buy the cloud-based CPU time). For example, such cloud resources were used in  this
paper.
But uses of cloud-based resources have several disadvantages:
\bi
\item 
Cloud computing environments are extensively monetized 
so the total financial cost of tuning can be prohibitive.
\item 
That CPU time is wasted if there is a faster and more effective way.
\ei
It turns out that the last point is indeed the case-- 
at least for tuning  defect predictors learned
from static code attributes. The case study of this paper compares
two tuning methods:
the {\em grid search}  as used by 
Tantithamthavorn et al.~\cite{tantithamthavorn2016automated}
and   {\em differential evolution} as used by Fu et al.~\cite{fu16}.
B  oth paper investigate the impacts of tunings on defect predictors and
find that parameter tuning can improve the learner performance by different methods.
This study makes the following contributions:

\begin{table*}[!hpt]
\renewcommand{\baselinestretch}{0.9}

\caption{OO measures used in our defect data sets.}\label{tab:ck}
\begin{center}
{\scriptsize
\begin{tabular}{l|l|p{5.0in}}
\rowcolor{lightgray}
\hline
Metric & Name & Description \\ \hline
amc & average method complexity & Number of JAVA byte codes\\\hline
avg\_cc & average McCabe & Average McCabe's cyclomatic complexity seen
in class\\\hline
ca & afferent couplings & How many other classes use the specific
class. \\\hline
cam & cohesion amongst classes & Summation of number of different
types of method parameters in every method divided by a multiplication
of number of different method parameter types in whole class and
number of methods. \\\hline
cbm &coupling between methods &  Total number of new/redefined methods
to which all the inherited methods are coupled\\\hline
cbo & coupling between objects & Increased when the methods of one
class access services of another.\\\hline
ce & efferent couplings & How many other classes is used by the
specific class. \\\hline
dam & data access & Ratio of  private (protected)
attributes to   total   attributes\\\hline
dit & depth of inheritance tree &  It's defined as “the maximum length from the node to the root of the tree”\\\hline
ic & inheritance coupling &  Number of parent classes to which a given
class is coupled (includes counts of methods and variables inherited)
\\\hline
lcom & lack of cohesion in methods &Number of pairs of methods that do
not share a reference to an instance variable.\\\hline
locm3 & another lack of cohesion measure & If $m,a$ are  the number of
$methods,attributes$
in a class number and $\mu(a)$  is the number of methods accessing an
attribute, 
then
$lcom3=((\frac{1}{a} \sum_j^a \mu(a_j)) - m)/ (1-m)$.
\\\hline
loc & lines of code & Total lines of code in this file or package.\\\hline
max\_cc & Maximum McCabe & maximum McCabe's cyclomatic complexity seen
in class\\\hline
mfa & functional abstraction & Number of methods inherited by a class
plus number of methods accessible by member methods of the
class\\\hline
moa &  aggregation &  Count of the number of data declarations (class
fields) whose types are user defined classes\\\hline
noc &  number of children & Number of direct descendants (subclasses) for each class\\\hline
npm & number of public methods & npm metric simply counts all the methods in a class that are declared as public. \\\hline
rfc & response for a class &Number of  methods invoked in response to
a message to the object.\\\hline
wmc & weighted methods per class & A class with more member functions than its peers is considered to be more complex and therefore more error prone \\\hline

defect & defect & Boolean: where defects found in post-release bug-tracking systems. \\\hline
\end{tabular}
}
\end{center}

\end{table*}

\bi
\item
To the best of our knowledge, this is the first such comparison
of  these two techniques for the purposes
of tuning defect predictors;
\item 
We show that differential evolution is {\em just as effective
as grid search for improving defect predictors}, while  
being {\em one to two orders of magnitude faster};
\item
Hence, we offer a strong recommendation to use evolutionary algorithms(like DE) for 
tuning defect predictors;
\item
Lastly,  we
 propose a prediction method that would allow future analysts
to match the optimization method to the data at hand.
\ei
The rest of this paper is structured as follows. Section 2
describes  defect predictors,  how they can
be generated by data miners, and how tuning can affect that learning process.
Section 3 defines and  compares differential evolution and grid search
for that tuning process. This section's case study
shows  that grid search is far slower, yet no more effective,
than differential evolution. Section 4 is a
discussion on why DE is more effective and faster than grid search
by showing how {\em intrinsic dimensionality} can be used 
as a predictor for tuning problems that might be 
suitable for random search methods such as differential evolution.

While the specific conclusion of this paper relates to defect prediction,
this work has broader implications across data science.
Many researchers in SE explore hyper-parameter tuning~\cite{Arcur11,Jia15,panichella2013effectively,cora10,minku2013analysis,song2013impact,lessmann2008benchmarking,tantithamthavorn2016automated,fu16,Wang:Harman13}
(for details, see Section 2.3).
Within that group, grid search is the common method for exploring
different tuning parameters. Such studies can be tediously slow, often requiring days to weeks
of CPU time. The success of  differential
evolution to tune defect predictors raises the possibility
that  tuning research could be greatly accelerated by a better
selection of tuning algorithms. This could be a very productive avenue for
future research.

\subsection{Case Studies, Not Experiments}
In terms of SE experimental theory this paper
is a {\em case study} paper and not an {\em experiment}.
Runeson and H\"ost~\cite{Runeson2009} reserve the term ``case
study'' to refers to some act which includes observing or
changing the conditions under which humans perform
some software engineering action. In the age of data mining
and model-based optimization, this definition should be
further extended to include case studies exploring how data
collected from different projects can be generalized across
multiple projects. As Cruzes et al. comment ``Choosing a suitable method for synthesizing evidence across a
set of case studies is not straightforward... limited access to raw
data may limit the ability to fully understand and synthesize
studies''~\cite{Cruzes11}.

In this age of model-based reasoning,
that limited information can be synthesized using a data mining. 
Such tools are automatic to use which means it is issue to use them automatically
and incorrectly. For example, as shown in this paper, it is not good enough
to use these learners with their default tunings since what is learned
from data miner can be greatly changed and improved by tuning. That is, tuning is essential.

The problem with tuning is that, if done using standard methods it can be needlessly computationally expensive. Hence, the goal of this paper is to comment on how
quickly tuning can be applied {\em without} incurring excessive CPU costs.
 
\section{Background}
\subsection{Defect Prediction}

Human programmers are clever, but flawed. Coding adds functionality, but also defects. 
Hence, software sometimes crashes(perhaps at the most awkward or dangerous moment) or delivers the wrong functionality.    

Since programming inherently introduces defects into programs,
it is important to test them before they are used. Testing is expensive. 
According to Lowry et al.~\cite{LowryBK98}, software assessment budgets are
finite while assessment effectiveness increases exponentially with assessment effort.
Exponential costs quickly exhaust finite resources 
so standard practice is to apply the  best  available methods 
only on code sections that seem most critical. 
Any method that focuses on parts of the code can miss defects in other areas 
so some sampling policy should be used to explore the rest of the system.
This sampling policy will always be incomplete, 
but it is the only option when resources prevent a complete assessment of everything.

\renewcommand\arraystretch{1.2}
\begin{figure*}[htp!]
\small
  \centering
	\begin{tabular}{|c|c|c|c|l|}
	\cline{1-5}
	\begin{tabular}[c]{@{}c@{}}Learner Name\end{tabular} & Parameters & Default &\begin{tabular}[c]{@{}c@{}}Tuning\\ Range\end{tabular}& 
\multicolumn{1}{c|}{Description} \\ \hline
 \multirow{5}{*}{CART} & threshold & 0.5 &[0,1]& The value to determine defective or not. \\ \cline{2-5} 
	 & max\_feature & None &[0.01,1]& The number of features to consider when looking for the best 
split. \\ \cline{2-5} 

	 & min\_sample\_split & 2 &[2,20]& The minimum number of samples required to split an 
internal node. \\ \cline{2-5} 
	 & min\_samples\_leaf & 1 & [1,20]&The minimum number of samples required to be at a leaf 
node. \\ \cline{2-5} 
     & max\_depth & None & [1, 50]& The maximum depth of the tree. \\
     \cline{1-5}  
       \multirow{5}{*}{\begin{tabular}[c]{@{}c@{}}Random \\ Forests\end{tabular}}  & threshold & 0.5 & [0.01,1] & The value to determine defective or not. \\ 
\cline{2-5} 
	 & max\_feature & None &[0.01,1]& The number of features to consider when looking for the best 
split. \\ \cline{2-5} 
	 & max\_leaf\_nodes & None &[1,50]& Grow trees with max\_leaf\_nodes in best-first fashion. \\ \cline{2-5} 
	 & min\_sample\_split & 2 &[2,20]& The minimum number of samples required to split an 
internal node. \\ \cline{2-5} 
	 & min\_samples\_leaf & 1 &[1,20]&The minimum number of samples required to be at a leaf 
node. \\ \cline{2-5} 
	 &  n\_estimators & $100$  & [50,150]&The number of trees in the forest.\\ \cline{2-5}
	 \hline 
	\end{tabular}
    \caption {List of parameters tuned by this paper.}
\label{fig:parameters}
\end{figure*}

One of sampling  policies  is  defect  predictors learned from static code attributes. 
Given software described in  the attributes like \fig{ck}, 
data  miners  can  infer where probability of software defects is most likely.
This is useful since static code attributes can be automatically collected, even for very large systems.
Other methods, like  manual code reviews, are slower and  more labor-intensive~\cite{rakitin01}. 
There are increasing number of work to build defect predictors based on static code attributes~\cite{krishna2016too,nam2015heterogeneous,tan2015online}.
and it can localize 70\% (or more)
of the defects in code~\cite{me07b}. They also perform well compared to certain
widely-used automatic methods.
Rahman et al.~\cite{rahman14:icse} compared
(a) static code analysis tools FindBugs, Jlint, and Pmd and (b)
static code defect predictors.
They found no significant differences in the cost-effectiveness
of these approaches. This is interesting
since static code defect prediction can be quickly adapted to new languages by building lightweight
parsers that find  information like \fig{ck}. 
The same is not true for static code analyzers which need 
extensive modification before they can be used on new
languages.

\subsection{Data Mining}

Data miners produce summaries of data.
Data miners are efficient since they employ various 
heuristics to reduce their search space for finding summaries.

For examples of how these heuristics control data miners, consider the 
CART and  Random Forests tree learners.
These algorithms  divide a data set, then recursively split on
on each nodes until some stop criterion is satisfied.
In the case of building defect predictors, these learners
reflect on the number of issue reports $d_i$ raised
for each class in a software system where the issue counts are converted
into binary ``yes/no'' decisions via \eq{yesno}, where T is a threshold number.

\begin{equation}\label{eq:yesno}
\mathit{inspect}= \left\{
\begin{array}{ll}
d_i \ge T \rightarrow \mathit{Yes}\\
d_i <   T \rightarrow \mathit{No} ,
\end{array}\right.
\end{equation}
For the specific implementation in Scikit-learn\cite{scikit-learn},
the splitting process is controlled by numerous tuning parameters
listed in \fig{parameters}, where the  default  parameters for CART and Random Forest are set by the Scikit-learn authors except for {\em n\_estimators}, as recommended by Witten et al.~\cite{Witten2011}, we used a forest of 100 trees as default instead of 10.
If data contains more than {\em min\_sample\_split}, then a split is attempted.
On the other hand, if a split contains no more than {\em min\_samples\_leaf}, then the recursion stops.  

These learners use different techniques to explore the splits:
\bi
\item
CART finds the attributes of the dataset whose ranges contain rows(samples of data) with least variance in the number
of defects: if an attribute ranges $r_i$ is found in 
$n_i$ rows, each with a  defect count variance of $v_i$, then CART seeks the attributes
whose ranges minimizes $\sum_i \left(\sqrt{v_i}\times n_i/(\sum_i n_i)\right)$.
\item
Random Forests divides data like CART then  builds number of $F$ trees($F>1$),
each time using some random subset of
the attributes. 
\ei
Note that some tuning parameters are learner specific.
{\em max\_feature} is used by
CART and Random Forests to select the number of attributes
used to build one tree.
CART's default is to use all the attributes while 
Random Forests usually selects the square root of the number
of attributes.
Also, {\em max\_leaf\_nodes} is the upper bound on leaf notes generated in a Random Forests.
Lastly, {\em max\_depth} is the upper bound on the depth of the CART tree.

\subsection{Parameter Tuning}

Parameter tuning for evolutionary algorithms
was studied by Arcuri \& Fraser and presented at  
SSBSE'11~\cite{Arcur11}. Using the grid search method (described below),
they found that different parameter settings for evolutionary
programs cause
very large variance in their performance.
Also, while parameter settings perform relatively well, they are
far from optimal on particular problem instances.

Tuning is now explored in many parts of the SE research literature.
Apart for defect prediction, tuning is used in
the hyper-parameter optimization literature exploring better
combinatorial search methods for software testing~\cite{Jia15}
or the use of genetic algorithms to explore 9.3 million
different configurations for clone detection algorithms~\cite{Wang:Harman13}.

Other researchers explore the effects of parameter tuning
on topic modeling~\cite{panichella2013effectively} (which is a
text mining technique). Like Arcuri \& Fraser, that work showed
that the performance of the LDA topic modelling algorithm
was greatly effected by the choice of four parameters that
configure LDA. Furthermore, Agrawal et.al~\cite{amrit16} demonstrate that more stable
topics can be generated by tuning LDA parameters using differential
evolution algorithm. 
 
Tuning  is also used for  software effort
estimation; e.g. using tabu search for tuning SVM~\cite{cora10};
or genetic algorithms for tuning ensembles~\cite{minku2013analysis};
or as an exploration tool for   checking if
parameter settings affect the performance of effort estimators 
(and what learning machines are more sensitive to their parameters)~\cite{song2013impact}. 
The latter study explored Random Forests, kth-nearest neighbor methods,
MLPs, and bagging. This was another grid search paper 
that explored a range of tunings parameters, divided into 5 bins.

Our focus is on defect prediction and for this task,
Lessmann et al. used
grid search to tune parameters as part of their extensive
analysis of different algorithms for defect prediction~\cite{lessmann2008benchmarking}.
Strangely, they only tuned a small set of their learners while for most
of them, they used the default settings. This is an observation we cannot
explain but our conjecture is that the overall cost of their grid search
tuning was so expensive that they restricted it to just the hardest choices.
To be fair to Lessmann et al.,  we note that the general
trend in SE data science papers is to explore defect prediction
without tuning the parameters; e.g.~\cite{me07b}.

Two exceptions to this rule are recent work by Tantithamthavorn et al.~\cite{tantithamthavorn2016automated} and Fu et al.~\cite{fu16}: the former used grid search
while the latter used differential evolution (both these techniques are detailed,
below). These two teams worked separately 
using completely different scripts (written in ``R'' or in Python).
Yet for tuning defect predictors, both groups reported the same results:
\bi
\item
Across a range of performance measures (AUC, precision, recall, F-measure),
tuning rarely makes performance worse;
\item Tuning offers a median improvement of 5\% to 15\%;
\item 
For a third of data sets exploration, tuning can result
in performance improvements of 30\% to 50\%.
\ei
Also, in a result that echoes one of the conclusions of  Arcuri \& Fraser,
Fu et al. report that different data sets require different tunings.

What was different between Fu et al. and Tantithamthavorn et al.
was the computational costs of the two studies.
Fu et al. used a single desktop machine and all their runs terminated in 
2 hours for one tuning goal.
On the other hand, the grid search of Tantithamthavorn et al.
used 43 high
performance computing machines with  
24 hyper-threads times 43 machines = 1,032 hyper-threads. Their total
runtime were not reported-- but as shown below, such tuning with grid
search can take over a day just
to learn one defect predictor.

\section{Comparing Grid Search and Differential Evolution}

Tantithamthavorn et al.~\cite{tantithamthavorn2016automated} and Fu et al.~\cite{fu16} use different
 methods to tune defect predictors. Neither offer a comparison of
 their preferred tuning method to any other. This section offers such a case study
 result: specifically, a comparison of grid search  and different evolution  
 for tuning defect predictors. 
\subsection{Algorithms}

 {\em Grid search} is simply picking a set of values for each configuration parameter and 
 evaluating all the combinations of these values, and then return the best one
 as the final optimal result, which can be simply implemented by nested for-loops.
 For example, for Naive Bayes, two loops might explore different
 values from the Laplace and M-estimator while a third
 loop might explore what happens when numeric 
 values are divided into, say,  $2 \le b \le 10$ bins.

Bergstra and  Bengio~\cite{Bergstra2012} comment on the popularity of
grid search:
\bi
\item
Such a simple search gives researchers some degree of insight into it;
\item
There is little technical overhead or barrier to its implementation;
\item As to automating grid search, it is  simple to implement and parallelization is trivial;
\item According to Bergstra and  Bengio~\cite{Bergstra2012},
grid search (on a  compute cluster) can find  better    tunings than 
sequential optimization (in the same amount of time).
\ei
Since it's easy to understand and implement and, to some extent, also has good
performance, grid search has been available in most popular data mining and machine
learning tools, like {\it caret}~\cite{kuhn2014caret} package in
the $R$ and {\it GridSearchCV} module in Scikit-Learn.

{\em Differential evolution}
is included in many optimization toolkits, like {\it JMetal} in Java~\cite{durillo2011jmetal}. But given its
implementation simplicity, it is often written from
scratch using the researcher's preferred scripting language.
 Differential evolution just randomly picks three different vectors  
$B,C,D$ from a list called $F$ (the {\em frontier}) for each parent vector A in $F$ ~\cite{storn1997differential}. 
Each pick generates a new
vector $E$ (which replaces $A$ if  it scores better).
$E$ is generated as follows:
\begin{equation} \label{eq:de}
  \forall i \in A,  E_i=
    \begin{cases}
      B_i + f*(C_i - D_i)& \mathit{if}\;  \mathcal{R} < \mathit{cr}  \\
      A_i&   \mathit{otherwise}\\ 
    \end{cases}
\end{equation}
where $0 \le \mathcal{R} \le 1$ is a random number,
and $f,cr$ are constants (following
Storn et al.~\cite{storn1997differential}, we use $cr=0.3$ and $f=0.75$).
Also, one $A_i$ value (picked at
random)
is moved to $E_i$ to ensure that $E$ has at least one
unchanged part of an existing vector.

As a sanity check, we also provide {\em random search} as a third optimizer to tune the parameters. 
Random search is nothing but randomly generate a 
set of different candidate parameters, and always
evaluate them against the current ``best'' one. 
If better, then it will replace the ``best'' one.
The process will repeat until the stop condition meets.
In this case study, we set maximum iterations for random search the same as
median number of evaluations in DE. The parameter will be randomly generated from the 
same tuning range as in \fig{parameters}.

\begin{figure*}[!htp]
\renewcommand{\baselinestretch}{0.8}
\small
\centering
  \begin{tabular*}{0.97\textwidth}{l |c c c c c c c c c } \hline
  Dataset &antV0&antV1&antV2&camelV0&camelV1&ivy&jeditV0&jeditV1&jeditV2
\\\hline
  training (release $i$) &20 / 125 &40 / 178 &32 / 293 &13 / 339 &216 / 608 &63 / 111 &90 / 272 &75 / 306 &79 / 312
\\  tuning (release $i+1$) &40 / 178 &32 / 293 &92 / 351 &216 / 608 &145 / 872 &16 / 241 &75 / 306 &79 / 312 &48 / 367
\\  testing (release $i+2$) &32 / 293 &92 / 351 &166 / 745 &145 / 872 &188 / 965 &40 / 352 &79 / 312 &48 / 367 &11 / 492
\\ \hline
  Dataset &log4j&lucene&poiV0&poiV1&synapse&velocity&xercesV0&xercesV1
\\\hline
  training (release $i$) &34 / 135 &91 / 195 &141 / 237 &37 / 314 &16 / 157 &147 / 196 &77 / 162 &71 / 440
\\  tuning (release $i+1$)  &37 / 109 &144 / 247 &37 / 314 &248 / 385 &60 / 222 &142 / 214 &71 / 440 &69 / 453
\\  testing(release $i+2$)  &189 / 205 &203 / 340 &248 / 385 &281 / 442 &86 / 256 &78 / 229 &69 / 453 &437 / 588
\\  \end{tabular*}

   \caption{Data used in this case study. Fractions denote
   {\em defects / total}.
   E.g., the top left data set has 20 defective classes out of 125 total.
   See  explanation in section~\ref{explain}.
   }\label{fig:data1}
\end{figure*}

Grid search is much slower than DE since DE explores
fewer options. Grid search's execution of 
  $X$ loops exploring $N$ options 
takes time $O(N^X)$, where $X$ is the number of parameters being tuned. Hence:
\bi
\item
Tantithamthavorn et al. required 1000s of hyperthreads to complete
their study in less than a day~\cite{tantithamthavorn2016automated}. 
\item
The grid search of Arcuri \& Fraser~\cite{Arcur11} took weeks to terminate,
even using a large computer cluster. 
\item
In the following study,
our  grid search times  took 54 days of
total CPU time.
\ei
On the other hand, DE's runtime 
are much faster since they are linear on the size of the frontiers, i.e. $O(|F|)$.

\subsection{Case Studies}
In order to understand the relative merits of grid search versus
differential evolution for defect prediction, we performed the following study.

\subsubsection{Data Miners}

This study uses Random Forests and CART, for the following reason.
Firstly, they were two of the learners studied by Tantithamthavorn et al.
Secondly, they are interesting learners in that they represent
two ends of a performance spectrum for defect predictors.
CART and Random Forests were mentioned in a recent IEEE TSE paper by Lessmann et al.~\cite{lessmann2008benchmarking} that compared 22 learners for defect prediction. That study ranked CART worst and Random Forests as best.  In a demonstration of the impact of tuning, 
Fu et al. showed that they could 
refute the conclusions of Lessmann et al. in the sense that, after tuning, CART performs just as well as Random Forests.

\subsubsection{Tuning Parameters}
Our DE and grid search explored the parameter space of
\fig{parameters}. Specifically, since Tantithamthavorn et al. divide each tuning range into 
5 bins (if applicable), we also use the same policy here. For example, we 
pick values$[50, 75, 100, 125, 150]$ for $n\_estimators$. Other parameters grid will generate
in the same way. As to why we used the ``Tuning Range'' shown in \fig{parameters}, and not some other ranges, we note that (1) those ranges included the defaults; (2) the results shown below show that by exploring those ranges, we achieved large gains in the performance of our defect predictors. This is not to say that larger tuning ranges might not result in greater improvements.

\subsubsection{Data}\label{explain}
Our defect data, shown in \fig{data1} comes from PROMISE 
repository\footnote{http://openscience.us/repo}.  This data
pertains to 
open source Java systems defined in terms of \fig{ck}:  
{\it ant}, {\it camel}, {\it ivy}, {\it jedit}, {\it log4j}, {\it lucene},
{\it poi}, {\it synapse}, {\it velocity} and {\it xerces}. 
We selected these data sets since they have  at least three  
consecutive releases  (where release $i+1$ was built after release $i$). 
This will allow us to build defect predictors based on the past data and then
predict (test) defects on future version projects, which will
be a more practical scenario.

More specifically, when tuning a learner:
\bi
\item
Release $i$ was used for training a learner with tunings generated by grid search or differential evolution.
\item
During the search,
each candidate has to be evaluated by some model,   which we build using
CART or Random Forests from
 release $i+1$. 
  \item After grid search or differential evolution terminated, we tested
  the tunings found by those methods on CART or Random Forests applied
  to release $i+2$.
  \item
  For comparison purposes, CART and Random Forests were also trained
  (with default tunings) on releases $i$ and $i+1$ then tested on release $i+2$.
\ei

\subsubsection{Optimization Goals:}
Our optimizers explore tuning improvements for precision and the F-measure,
defined as follows. Let $\{A, B, C, D\}$ denote the
true negatives, 
false negatives, 
false positives, and 
true positives
(respectively) found by a binary detector. 
Certain standard measures can be computed from
$A,B,C,D$, as shown below. Note that for $pf$, the {\em better} scores are {\em smaller}
while
for all other scores, the {\em better} scores are {\em larger}.

{\[
\begin{array}{ll}
pd=recall=&D/(B+D)\\
pf=&C/(A+C)\\ 
prec=precision=&D/(D+C) \\
F =&2*pd*prec/(pd + prec)
\end{array}
\]}

\begin{figure*}[ht]
\centering
\includegraphics[width=3.6in]{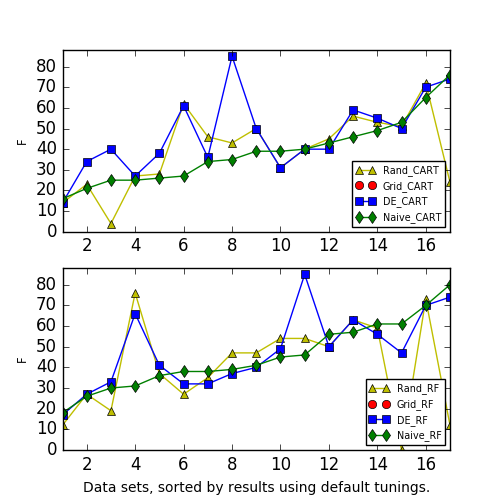}\includegraphics[width=3.6in]{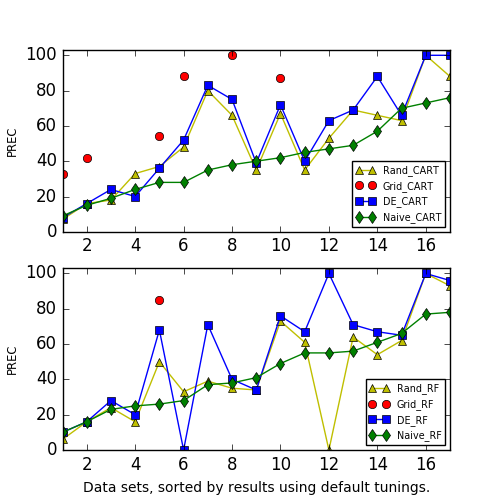}
\caption{Tuning to improve $F$ and $Precision$.  Median results from 20 repeats. 
F-measures results shown on left. Precision results shown on right.
\textcolor{blue}{{\bf Blue = DEs}},
\textcolor{OliveGreen}{{\bf Green = no tunings}},
\textcolor{Yellow}{\bf Yellow = random search}.
\textcolor{red}{{\bf Red= grid search}} results that are
better and \textbf{statistically} different to DE.}
\label{fig:fWei}
\end{figure*}

We do not explore all goals since some have trivial, but not useful, solutions.
For example,
when  we  tune  for recall, we can achieve near 100\% recall– but the cost of a
near 100\% false alarms. Similarly, when  we  tune  to minimize false alarms,
we can achieve near 0\% false alarms– but the cost of a near 0\% recall.

The lesson here
is that tuning for defect predictors needs some ``brake'' effect where
multiple goals are in contention (so one cannot be pushed
to some extreme value without being ``braked'' by the other). Precision's definition takes into accounts
not only the defective examples but also the none defective ones as well so
it has this brake effect. The same is true for
the $F$ measure (since it uses precision).

\subsubsection{20 Random Runs:}

All our studies were repeated $20$ times to check for the stability
of conclusions across different random biases. Initially, we planned for $30$ repeats
(appealing to the central limit theorem) but grid search proved to be so slow
that, for pragmatic reasons, we used $20$ repeats. {\it To be clear, the random
seed is different for each data set in each repeat, but it will be the same across learners
built by grid search and DE as well as random search for the same data set.}


The reason that we believe this is the right thing to do is
the search bias for a particular train/tune/test run is always the same. 
For the same train/tune/test data set, searching algorithms will start from
the same random seed.
With this approach, it is important to note that 
 different triplets have different seed values (so this
 case study does sample across a range of search biases).

\subsubsection{Statistical Tests}

For each data set, the results of grid search and DE were compared across the 20 repeats.
Statistical differences were tested by the Scott-Knott test~\cite{scott1974cluster} that used the Efron \& Tibshirani bootstrap procedure~\cite{efron93}
and the Vargha and Delaney A12 effect size test~\cite{Vargha00} to confirm that 
sub-cluster of the treatments are statistically
different by more than a small effect. We used these statistical  tests
since they
were recently endorsed in the SE literature by Mittas \& Angelis (for Scott-Knott)
in TSE'13~\cite{mittas13} and Acura \& Briand (for A12) at ICSE'11~\cite{arcuri11}.
 
\subsection{Results}\label{results}
In this section, we will present the results from the above designed case studies.
To better compare DE with grid search, we'd like to investigate the following questions :
\bi
\item Does tuning improve learners performance?
\item Is grid search statistically better than DE in terms of performance improvements (precision \& F-measure)?
\item Is DE a more efficient tuner than grid search in terms of cost(runtime)? 
\ei

\subsubsection{Performance Improvements}
The changes in {\em precision} and{\em F-Measure} using parameters chosen by DE and grid search are shown in \fig{fWei}. As a note, \fig{fWei} is generated by two separate case studies with tuning goals of {\em precision} and{\em F-Measure}, respectively.
In that figure:
\bi
\item
The \textcolor{blue}{{\bf blue squares}} show the DE results;
\item
The 
\textcolor{OliveGreen}{{\bf green diamonds}} show what happens we run a learner
using the default parameter settings;
\item 
The
\textcolor{Yellow}{\bf yellow triangle} show the random search results;
\item
The  
\textcolor{red}{{\bf red dots}} indicate where the grid search results
were statistically different and better (i.e. greater than) than DE.
\ei

Over all, from \fig{fWei}, we can see that random search and DE both improve learners performance in terms of precision and F-measure, respectively. For example, in this case study, a simple random search can improve precision scores for CART and Random Forests in $\frac{11}{17}$ and $\frac{7}{17}$ data sets, respectively. The similar pattern can be found when F is set as our tuning objective. This further supports the conclusion from Fu et.al.\cite{fu16} that parameter tuning is helpful and can't be ignored.

\begin{lesson}
  Tuning can improve learner performance in most cases; DE is better than random search in a couple of data sets.
\end{lesson}

The key feature of \fig{fWei} is that the red dots are in the minority;
i.e. there are very few cases where the grid search generate better results
than DE:
\bi
\item When tuning CART:
\bi
\item
For the F-measure, grid search was never better than DE;
\item For precision, grid does better than DE in $\frac{6}{17}$ cases
(less than half the time).
\ei
\item When tuning Random Forests:
\bi
\item For the F-measure, there is only one case where grid does better than DE;
\item For precision, grid does better than DE in $\frac{1}{17}$ data sets;
i.e. not very often.
\ei
\ei
Note that \fig{fWei} also contains results that echoes
conclusions
from   Arcuri \& Fraser  in
SSBSE'11~\cite{Arcur11}:
\bi
\item 
{\em Default  parameter settings perform relatively well.} Note that there exists several cases where the tuned
results (blue squares or red dots) are not much different to the results
from using the default parameters (the green diamonds). 
\item
{\em But they are
far from optimal on particular problem instances.} Note all the
results   where the red dots and blue squares are higher than green, particular in the results
where we are tuning for precision.
\ei

\begin{lesson}
  Differential evolution is just as effective as grid search for improving defect predictors.
\end{lesson}
\subsubsection{Runtime}

\begin{figure}[!ht]
\centering 
\includegraphics[width=3.45in]{pic/runtime.png}
\caption{Time(in seconds)required to run the 20 repeats of
tunings to generate \fig{fWei}. Left-hand-side plot shows
raw runtime; Right-hand-side plot shows
runtime releative to just running the learners using their
default tunings.  
}\label{fig:time}
\end{figure}

\noindent
\fig{time} shows the runtime costs of grid search, random search and  differential evolution over 17 data sets including both F-measure and Precision as optimzation objectives:
\bi
\item
The \textcolor{orange}{{\bf orange plot}} shows the raw runtime (in seconds)
of our entire trial.
In that plot, 
\bi
\item
CART is faster than Random Forests: 20 repeats
of a grid search
of the latter required 54 days of CPU time;
\item
DE is faster than grid  search:
$10^3$ or $10^4$ seconds for DE vs $10^5$ or $10^6$ seconds for grid.
\item
DE runs as fast as random search for both CART and RF: there's no significant difference.
\ei
\item
The \textcolor{blue}{{\bf blue plot}}
shows that the same information but in ratio form. 
\bi
\item
Grid search  is 1,000s to 10,000s times slower than
just running the default learners 
\item
While DE adds a factor of  $10^2$ to the default(untuned) runtime cost.
\ei
\ei

\begin{lesson}
  Differential evolution runs much faster than grid search.
\end{lesson}

In summary:
\bi
\item Both DE and random search as well as grid search can improve precision and the F-measure.
\item
Compared to DE, grid search runs far too long for too little additional
benefit.
\item
Both DE and random search require the same amount of runtime. But for some cases,
DE has better performance than random search.
\item
There are many cases where DE out-performs grid search.

\ei

\section{Discussion}

\subsection{Why does DE perform better than grid search?}

How to explain the suprising success of DE over grid search?
Surely a thorough investigation of all options (via grid search)
must do better than a partial exploration of just a few options (via
DEs).

It turns out that grid search is not a thorough exploration of all options.
Rather,  it {\em jumps} through different parameter settings
between some {\em min} and {\em max} value of pre-defined tuning range.
If the best  options lie in between these jumps, then grid search will {\em skip} the
critical tuning values. That means, the selected grid points will
finally determine what kind of tunings we can get and good 
tunings require a lot of expert knowledge.

Note that DE is less prone to {\em  skip} since, as shown
in \eq{de}, tuning values are adjusted by some random amount that is
the difference between two randomly selected vectors. Further, if
that process results in a better candidate, then this new randomly
generated value might be used as the start point of a subsequent
random selection of data. Hence DE is more likely than grid search
to ``fill in the gaps'' between an initially selected values.

Another important difference between DE and grid search is the 
  nature of their searches:
\bi
\item
All the {\it grid points} in the pre-defined grids are independently 
evaluated. 
This is useful since it makes grid search
highly suited for parallelism (just
run some of the loops on different processors).
That said, this independence has a drawback:  
 any lessons learned midway by grid search
{\em cannot affect(improve)} the inferences made in the remaining runs.
\item 
DE's {\it candidates} (equivalent to grid points) do ``transfer knowledge'' 
to candidates in the new generation.
Since DE is an evolutionary algorithms, the better candidates
will be inherited by the following generations. That said,
DE's discoveries of better vectors
accumulate in the frontier-- which means  
new solutions(candidates) are being continually built from increasingly
better solutions cached in the frontier. 
That is, 
lessons learned midway through a DE run 
{\em can improve} the inferences made in the remaining runs.
\ei
Bergstra and Benigo~\cite{Bergstra2012}  
offer a more formal analysis for why random searches
(like DEs) can do better than grid search. They comment that grid search
will be expected to fail if the region containing the useful tunings
is very small. In such a search space:
\bi
\item
Grid search can waste much time exploring an irrelevant part of the space.
\item
Grid search's effectiveness is limited by the  {\em curse of dimensionality}.
\ei
 Bergstra and Benigo reasons for the second point are as follows.
They compared  deep belief networks configured by a thoughtful combination of
manual search and grid search, and purely random search over the same 32-dimensional configuration
space. They found found statistically equal performance on four of seven data sets, and superior performance
on one of seven.
A Gaussian process analysis of their systems
revealed that for most data sets only a few of the tuning  really matter,
but that different hyper-parameters are important on different data sets. 
They  comment that a grid
with sufficient granularity to tune for all data sets must consequently be
inefficient for each individual data set because of the curse of dimensionality: the number of wasted
grid search trials is exponential in the number of search dimensions that turn out to be irrelevant for
a particular data set. 
Bergstra and Benigo add:
\begin{quote}{\em ... in contrast, random search thrives on low effective dimensionality. Random
search has the same efficiency in the relevant subspace as if it had been used to search only the
relevant dimensions.}\end{quote} 
 
 Our previous results in Section \ref{results} also verified Bergstra and Benigo's conclusion that random search is
 much better than grid search for exploring defect prediction data space, where random search is as good as DE in most data sets. But grid search rarely outperformed DE and random search in terms of performance scores(F-measure and Precision). However, grid search just waste a lot of time to explore unnecessary space.

\subsection{When (not) to use DE?}
How might we assess the external validity of the above results?
Is it possible to build some predictor when DEs might and might not work well? 

To explore these questions
we use Bergstra and Benigo's comments 
to define the conditions when we would expect DEs to work better
than grid search for defect prediction. According to the argument above, DE
works well for tuning since:
\bi
\item DE tends to favor the small number of dimensions relevant to tuning;
\item The space of tunings for defect predictors is inherently low dimensional.
\ei
In defence of the  first point, recall that
\eq{de} says that DE repeatedly
compares an existing tuning $A$ against another candidate $E$ that
is constructed by taking a small step between three other candidates $B,C,D$.
DE runs over a list of old candidates, $n$ times. For $n>1$, the invariant
 is that   members of that list are  not inferior
to  at least one other example.  If a new candidate $E$ is
created that is orthogonal to the relevant dimensions, it will be no
better than the candidates $B,C,D$ it was created from. Hence, the invariant
for any successful $E$ replacement of $A$ is that it has moved over the relevant
dimensions.

As to the second point about the low dimensional nature of tuning
defect predictors, we first assume that {\em the dimensionality of the tuning
problem is linked to the dimensionality of the data explored by
the learners}. Our argument for this assumption is (1)~learners like CART and Random Forest
  divide the data into regions with similar properties (specifically,
those with and without defects); (2)~when we tune those  learners, we are constraining how they make those divisions
over that data.

\begin{figure*}[h]
\begin{center}
\includegraphics[width=5in]{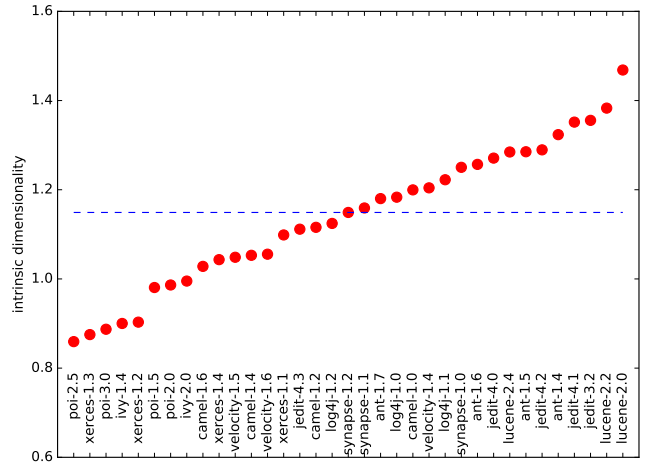}
\end{center}
\caption{Intrinsic dimensionality of our data. }\label{fig:id}
\end{figure*}
Given that assumption, exploring the space of tunings for defect predictors 
really means {\em exploring the dimensionality of defect prediction
data}. Two studies strongly suggest that this data is inherently low-dimensionality.
Papakroni~\cite{papa13} combined instance selection and attributes pruning for
defect prediction. Using some information theory,
he was able to prune 75\% of the attributes of \fig{ck} as uninformative.
He then clustered the remaining data,  replacing
each cluster with one centroid. This two-phase pruning
procedure generated small data sets with, e.g., 24 columns~(attributes) and 800
rows~(instances) to a table  of 6 columns and 20 rows.  
To test the efficacy of that reduced space,
Papakroni built defect predictors by  
extrapolating between the two nearest centroids in the reduced space, for each test case. Papakroni found that those  estimates from that small
\begin{wrapfigure}{r}{1.7in}
\includegraphics[width=1.7in]{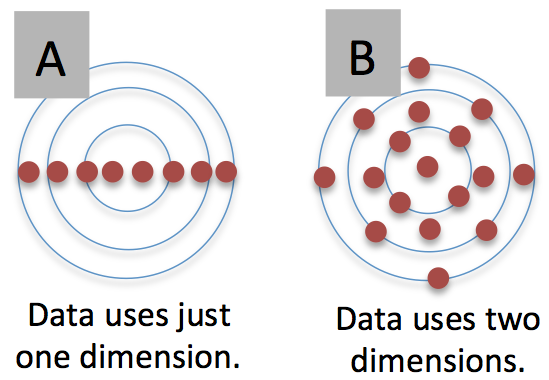}
\end{wrapfigure}
 space worked just
as well
as those generated by
state-of-the-art learners (Random Forests and
Naive Bayes) using all the data~\cite{papa13}. That is, according to
  Papakroni,
the signal in these data sets can be found in a handful
of attributes and a few dozen instances.

In order to formalize the findings of the
Papakroni study,
Provence~\cite{province15} explored the  {\em intrinsic 
dimensionality} of defect data sets.
{\em Intrinsic dimensionality}
measures the number of dimensions $m$ used by data within an $n$ dimensional
space. 
For example, the ``B'' data shown at right  spreads   over   a $n=2$ dimensional space.
but the ``A'' data  does not use all  the available dimensions.
Hence, the intrinsic dimensionality of the ``A'' data is  $m=1$.
 
Like Provence, we use correlation dimension to calculate the intrinsic dimensionality of the datasets. Euclidean distance is used to compute the distance between the independent decisions $d$ within each candidate solution; all $d_i$ values are normalized by $max-min$. Next, we use the distance measure as part of the
{\em correlation dimension} defined by
by Grassberger and Procaccia\cite{grassberger1983measuring}.
This  correlation dimension of a   data set with $k$ items
is found by  plotting the number of items found at distance within radius \emph{r} from any other item against $r$ (where $r$
is actually a distance, as defined in the last paragraph). Then we
normalize this by the number of connections between $k$ items
to find the expected number of neighbors at distance \emph{r} is
$C(r)=\frac{2}{k(k-1)}\sum_{i=1}^n\sum_{j=i+1}^n 1\{{\lVert}x_i,x_j{\rVert}<r\}$.



Given a data set with $k$ items and min, max distance of $r_{min}$ and $r_{max}$,
we estimate the intrinsic dimensionality as the mean value 
of the slope  of $\ln{(C(r))}$ vs $\ln{(r)}$ by evaluating $C(r)$ for $r$ in $\{r_0, ..., r_n\}$, such that $\{r_0, ..., r_n\}$ is sufficient for a good estimation of slope, and \mbox{$r_n<<r_{max}$}.  

\fig{id} shows the intrinsic dimensions for the data sets used in this
study. Note the low intrinsic dimensionality (median value, shown as the dashed line,
nearly 1.2). By way of comparison,
the intrinsic dimensions reported 
by other researchers in their data sets (not from SE) 
is often much larger; e.g.
5 to 10, or more; see~\cite{levina04}.   

From all this, Provence
concluded that the effects reported by Papakroni were due to  the underlying low
dimensionality of the data. Extending his result, we conjecture that our conclusion
(that DEs do better than grid search for tuning data miners) are externally valid
when the data miners are exploring data with low intrinsic dimensionality.

\section{Conclusion and Future Work}

When the control parameters of data miners are tuned
to the local data, the performance of the resulting predictions
can be greatly increased. Such tuning are   very computationally 
expensive, especially when done with grid search.
For some software engineering tasks, it is possible
to avoid those very long runtime.
When tuning defect predictors learned from static code attributes,
a   simple randomized evolutionary strategy (differential
evolution) runs one to two orders of magnitude faster
that grid search. Further the tunings found in this way
work as well, or better, than those found via grid search.
We explain this result using the   (1)~Bergstra and Benigo argument that 
random search works best in low-dimensional data sets and (2)~the empirical
results of Papakroni and Provence that defect data sets are very
low dimensional in nature.

\noindent
As to testing the external validity of this paper's argument, the next steps
  are clear:
\be
\item Sort data sets by how well
a simple evolutionary algorithm like DE can improve the performance
of data miners executing on that data;
\item 
Explore the difference in the {\em worst} and 
{\em best} end of that sort.
\item
  If the intrinsic
dimensionalities are very different at both ends of this sort,
then that would
support with the claims of this paper .
\item 
Else,  this paper's claims would not be supported
and we would need to seek other difference between the {\em best}
and {\em worst} data sets.
\ee
Note that we offer these four items as future
work, rather than reported results, since
so far all the defect data sets we have tried had responded best to DE. 
For a time, we did consider trying this with
artificially generated   data   (where we control how many dimensions were contained in the data). However, prior experience with using artificial data sets~\cite{me99q} suggested to us that such arguments
are not widely convincing since the issue is not how often an effect holds in {\em artificial} data, but how often in holds in {\em real} data. 
Hence, in future work, we will  look
further afield for radically different
(real world) data sets.


\section*{Acknowledgments}
The work has partially funded by a National Science Foundation CISE CCF award \#1506586.

\bibliographystyle{elsarticle-num}
\balance
\bibliography{main}

\begin{thebibliography}{10}
\expandafter\ifx\csname url\endcsname\relax
  \def\url#1{\texttt{#1}}\fi
\expandafter\ifx\csname urlprefix\endcsname\relax\def\urlprefix{URL }\fi
\expandafter\ifx\csname href\endcsname\relax
  \def\href#1#2{#2} \def\path#1{#1}\fi

\bibitem{Menzies:2013}
T.~Menzies, T.~Zimmermann, Software analytics: So what?, IEEE Softw. 30~(4)
  (2013) 31--37.

\bibitem{me07b}
T.~Menzies, J.~Greenwald, A.~Frank, Data mining static code attributes to learn
  defect predictors, IEEE Trans. Softw Eng. 33~(1) (2007) 2--13.

\bibitem{fu16}
W.~Fu, T.~Menzies, X.~Shen, Tuning for software analytics: Is it really
  necessary?, Information and Software Technology 76 (2016) 135--146.

\bibitem{brieman00}
L.~Breiman, Random forests, Machine learning 45~(1) (2001) 5--32.

\bibitem{Arcur11}
A.~Arcuri, G.~Fraser, On parameter tuning in search based software engineering,
  in: Proceedings of the Third International Conference on Search Based
  Software Engineering, SSBSE'11, Springer-Verlag, Berlin, Heidelberg, 2011,
  pp. 33--47.

\bibitem{Wang:Harman13}
T.~Wang, M.~Harman, Y.~Jia, J.~Krinke, Searching for better configurations: a
  rigorous approach to clone evaluation, in: Proceedings of the 2013 9th Joint
  Meeting on Foundations of Software Engineering, ACM, 2013, pp. 455--465.

\bibitem{tantithamthavorn2016automated}
C.~Tantithamthavorn, S.~McIntosh, A.~E. Hassan, K.~Matsumoto, Automated
  parameter optimization of classification techniques for defect prediction
  models, in: Proceedings of the 38th International Conference on Software
  Engineering, ACM, 2016, pp. 321--332.

\bibitem{Jia15}
Y.~Jia, M.~B. Cohen, M.~Harman, J.~Petke, Learning combinatorial interaction
  test generation strategies using hyperheuristic search, in: Proceedings of
  the 37th International Conference on Software Engineering-Volume 1, IEEE
  Press, 2015, pp. 540--550.

\bibitem{panichella2013effectively}
A.~Panichella, B.~Dit, R.~Oliveto, M.~Di~Penta, D.~Poshyvanyk, A.~De~Lucia, How
  to effectively use topic models for software engineering tasks? an approach
  based on genetic algorithms, in: Proceedings of the 2013 International
  Conference on Software Engineering, IEEE Press, 2013, pp. 522--531.

\bibitem{cora10}
A.~Corazza, S.~Di~Martino, F.~Ferrucci, C.~Gravino, F.~Sarro, E.~Mendes, How
  effective is tabu search to configure support vector regression for effort
  estimation?, in: Proceedings of the 6th international conference on
  predictive models in software engineering, ACM, 2010, p.~4.

\bibitem{minku2013analysis}
L.~L. Minku, X.~Yao, An analysis of multi-objective evolutionary algorithms for
  training ensemble models based on different performance measures in software
  effort estimation, in: Proceedings of the 9th international conference on
  predictive models in software engineering, ACM, 2013, p.~8.

\bibitem{song2013impact}
L.~Song, L.~L. Minku, X.~Yao, The impact of parameter tuning on software effort
  estimation using learning machines, in: Proceedings of the 9th international
  conference on predictive models in software engineering, ACM, 2013, p.~9.

\bibitem{lessmann2008benchmarking}
S.~Lessmann, B.~Baesens, C.~Mues, S.~Pietsch, Benchmarking classification
  models for software defect prediction: A proposed framework and novel
  findings, IEEE Trans. Softw Eng. 34~(4) (2008) 485--496.

\bibitem{LowryBK98}
M.~Lowry, M.~Boyd, D.~Kulkami, Towards a theory for integration of mathematical
  verification and empirical testing, in: Automated Software Engineering, 1998.
  Proceedings. 13th IEEE International Conference on, IEEE, 1998, pp. 322--331.

\bibitem{rakitin01}
S.~Rakitin, Software Verification and Validation for Practitioners and
  Managers, Second Edition, Artech House, 2001.

\bibitem{krishna2016too}
R.~Krishna, T.~Menzies, W.~Fu, Too much automation? the bellwether effect and
  its implications for transfer learning, in: Proceedings of the 31st IEEE/ACM
  International Conference on Automated Software Engineering, ACM, 2016, pp.
  122--131.

\bibitem{nam2015heterogeneous}
J.~Nam, S.~Kim, Heterogeneous defect prediction, in: Proceedings of the 2015
  10th joint meeting on foundations of software engineering, ACM, 2015, pp.
  508--519.

\bibitem{tan2015online}
M.~Tan, L.~Tan, S.~Dara, C.~Mayeux, Online defect prediction for imbalanced
  data, in: Proceedings of the 37th International Conference on Software
  Engineering-Volume 2, IEEE Press, 2015, pp. 99--108.

\bibitem{rahman14:icse}
F.~Rahman, S.~Khatri, E.~T. Barr, P.~Devanbu, Comparing static bug finders and
  statistical prediction, in: Proceedings of the 36th International Conference
  on Software Engineering, ACM, 2014, pp. 424--434.

\bibitem{scikit-learn}
F.~Pedregosa, G.~Varoquaux, A.~Gramfort, V.~Michel, B.~Thirion, O.~Grisel,
  M.~Blondel, P.~Prettenhofer, R.~Weiss, V.~Dubourg, J.~Vanderplas, A.~Passos,
  D.~Cournapeau, M.~Brucher, M.~Perrot, E.~Duchesnay, Scikit-learn: Machine
  learning in {P}ython, Journal of Machine Learning Research 12 (2011)
  2825--2830.

\bibitem{Witten2011}
I.~Witten, E.~Frank, {Data Mining, practical machine learning tools and
  techniques}, 3rd Edition, Morgan Kaufmann Pub, 2011.

\bibitem{amrit16}
A.~Agrawal, W.~Fu, T.~Menzies, What is wrong with topic modeling? (and how to
  fix it using search-based se), in: arXiv:1608.08176 [cs.SE], 2016, available
  from \url{http://arxiv.org/abs/1608.08176}.

\bibitem{Gao:2011}
K.~Gao, T.~M. Khoshgoftaar, H.~Wang, N.~Seliya, Choosing software metrics for
  defect prediction: An investigation on feature selection techniques, Softw.
  Pract. Exper. 41~(5) (2011) 579--606.

\bibitem{Moser:2008}
R.~Moser, W.~Pedrycz, G.~Succi, A comparative analysis of the efficiency of
  change metrics and static code attributes for defect prediction, in: ICSE
  '08, ACM, 2008, pp. 181--190.
\newblock \href {http://dx.doi.org/10.1145/1368088.1368114}
  {\path{doi:10.1145/1368088.1368114}}.

\bibitem{Elish2008649}
K.~O. Elish, M.~O. Elish, Predicting defect-prone software modules using
  support vector machines, Journal of Systems and Software 81~(5) (2008) 649 --
  660.

\bibitem{Bergstra2012}
J.~Bergstra, Y.~Bengio, {Random Search for Hyper-Parameter Optimization},
  Journal of Machine Learning Research 13 (2012) 281--305.

\bibitem{kuhn2014caret}
M.~Kuhn, J.~Wing, S.~Weston, A.~Williams, C.~Keefer, A.~Engelhardt, T.~Cooper,
  Z.~Mayer, R.~C. Team, M.~Benesty, et~al., caret: classification and
  regression training. r package version 6.0-24 (2014).

\bibitem{durillo2011jmetal}
J.~J. Durillo, A.~J. Nebro, jmetal: A java framework for multi-objective
  optimization, Advances in Engineering Software 42~(10) (2011) 760--771.

\bibitem{storn1997differential}
R.~Storn, K.~Price, Differential evolution--a simple and efficient heuristic
  for global optimization over continuous spaces, Journal of global
  optimization 11~(4) (1997) 341--359.

\bibitem{fu_wei_2017_344970}
W.~Fu, T.~Menzies, \href{https://doi.org/10.5281/zenodo.344970}{{Data set for
  ``Why is Differential Evolution Better than Grid Search for Tuning Defect
  Predictors?''}} (Mar. 2017).
\newblock \href {http://dx.doi.org/10.5281/zenodo.344970}
  {\path{doi:10.5281/zenodo.344970}}.
\newline\urlprefix\url{https://doi.org/10.5281/zenodo.344970}

\bibitem{rahman2012recalling}
F.~Rahman, D.~Posnett, P.~Devanbu, Recalling the imprecision of cross-project
  defect prediction, in: Proceedings of the ACM SIGSOFT 20th International
  Symposium on the Foundations of Software Engineering, ACM, 2012, p.~61.

\bibitem{song2011general}
Q.~Song, Z.~Jia, M.~Shepperd, S.~Ying, J.~Liu, A general software
  defect-proneness prediction framework, IEEE Trans. Softw Eng. 37~(3) (2011)
  356--370.

\bibitem{Mende:2010}
T.~Mende, R.~Koschke, Effort-aware defect prediction models, in: Proceedings of
  the 2010 14th European Conference on Software Maintenance and Reengineering,
  CSMR '10, IEEE Computer Society, Washington, DC, USA, 2010, pp. 107--116.
\newblock \href {http://dx.doi.org/10.1109/CSMR.2010.18}
  {\path{doi:10.1109/CSMR.2010.18}}.

\bibitem{yang2016defect}
J.~Yang, H.~Qian, Defect prediction on unlabeled datasets by using unsupervised
  clustering, in: High Performance Computing and Communications; IEEE 14th
  International Conference on Smart City; IEEE 2nd International Conference on
  Data Science and Systems (HPCC/SmartCity/DSS), 2016 IEEE 18th International
  Conference on, IEEE, 2016, pp. 465--472.

\bibitem{kamei2013large}
Y.~Kamei, E.~Shihab, B.~Adams, A.~E. Hassan, A.~Mockus, A.~Sinha, N.~Ubayashi,
  A large-scale empirical study of just-in-time quality assurance, IEEE
  Transactions on Software Engineering 39~(6) (2013) 757--773.

\bibitem{scott1974cluster}
A.~Scott, M.~Knott, A cluster analysis method for grouping means in the
  analysis of variance, Biometrics (1974) 507--512.

\bibitem{efron93}
B.~Efron, R.~J. Tibshirani, An introduction to the bootstrap, Mono. Stat. Appl.
  Probab., Chapman and Hall, London, 1993.

\bibitem{Vargha00}
A.~Vargha, H.~D. Delaney, A critique and improvement of the cl common language
  effect size statistics of mcgraw and wong, Journal of Educational and
  Behavioral Statistics 25~(2) (2000) 101--132.

\bibitem{mittas13}
N.~Mittas, L.~Angelis, Ranking and clustering software cost estimation models
  through a multiple comparisons algorithm, IEEE Trans. Software Eng. 39~(4)
  (2013) 537--551.

\bibitem{arcuri11}
A.~Arcuri, L.~Briand, A practical guide for using statistical tests to assess
  randomized algorithms in software engineering, in: Software Engineering
  (ICSE), 2011 33rd International Conference on, IEEE, 2011, pp. 1--10.

\bibitem{papa13}
V.~Papakroni, Data carving: Identifying and removing irrelevancies in the data,
  Master's thesis, Lane Department of Computer Science and Electrical
  Engineering, West Virginia Unviersity (2013).

\bibitem{province15}
B.~Province, Exploiting low dimensionality in software engineering, Master's
  thesis, CSEE, West Virginia University (2015).

\bibitem{grassberger1983measuring}
P.~Grassberger, I.~Procaccia, Measuring the strangeness of strange attractors,
  Physica D: Nonlinear Phenomena 9~(1) (1983) 189--208.

\bibitem{levina04}
E.~Levina, P.~J. Bickel, Maximum likelihood estimation of intrinsic dimension,
  in: Advances in neural information processing systems, 2004, pp. 777--784.

\bibitem{kocaguneli2012value}
E.~Kocaguneli, T.~Menzies, J.~W. Keung, On the value of ensemble effort
  estimation, IEEE Transactions on Software Engineering 38~(6) (2012)
  1403--1416.

\bibitem{me99q}
T.~Menzies, B.~Cukic, {When to Test Less}, IEEE Software 17~(5) (2000)
  107--112.

\end{thebibliography}


\begin{thebibliography}{10}
\expandafter\ifx\csname url\endcsname\relax
  \def\url#1{\texttt{#1}}\fi
\expandafter\ifx\csname urlprefix\endcsname\relax\def\urlprefix{URL }\fi
\expandafter\ifx\csname href\endcsname\relax
  \def\href#1#2{#2} \def\path#1{#1}\fi

\bibitem{Menzies:2013}
T.~Menzies, T.~Zimmermann, \href{http://dx.doi.org/10.1109/MS.2013.86}{Software
  analytics: So what?}, IEEE Softw. 30~(4) (2013) 31--37.
\newblock \href {http://dx.doi.org/10.1109/MS.2013.86}
  {\path{doi:10.1109/MS.2013.86}}.
\newline\urlprefix\url{http://dx.doi.org/10.1109/MS.2013.86}

\bibitem{me07b}
T.~Menzies, J.~Greenwald, A.~Frank, Data mining static code attributes to learn
  defect predictors, IEEE Trans. Softw Eng. 33~(1) (2007) 2--13, available from
  \url{http://menzies.us/pdf/06learnPredict.pdf}.

\bibitem{fu16}
W.~Fu, T.~Menzies, X.~Shen, Tuning for software analytics: is it really
  necessary?, Information and SoftwareTechnology (submitted)Read on-line at
  https://goo.gl/Jp5VIm.

\bibitem{brieman00}
L.~Breiman, Random forests, Machine learning 45~(1) (2001) 5--32.

\bibitem{Arcur11}
A.~Arcuri, G.~Fraser,
  \href{http://dl.acm.org/citation.cfm?id=204sa2243.2042252}{On parameter
  tuning in search based software engineering}, in: Proceedings of the Third
  International Conference on Search Based Software Engineering, SSBSE'11,
  Springer-Verlag, Berlin, Heidelberg, 2011, pp. 33--47.
\newline\urlprefix\url{http://dl.acm.org/citation.cfm?id=204sa2243.2042252}

\bibitem{Wang:Harman13}
T.~Wang, M.~Harman, Y.~Jia, J.~Krinke, Searching for better configurations: a
  rigorous approach to clone evaluation, in: Proceedings of the 2013 9th Joint
  Meeting on Foundations of Software Engineering, ACM, 2013, pp. 455--465.

\bibitem{tantithamthavorn2016automated}
C.~Tantithamthavorn, S.~McIntosh, A.~E. Hassan, K.~Matsumoto, Automated
  parameter optimization of classification techniques for defect prediction
  models, in: The International Conference on Software Engineering (ICSE),
  2016.

\bibitem{Jia15}
Y.~Jia, M.~B. Cohen, M.~Harman, J.~Petke, Learning combinatorial interaction
  test generation strategies using hyperheuristic search, in: Proceedings of
  the 37th International Conference on Software Engineering-Volume 1, IEEE
  Press, 2015, pp. 540--550.

\bibitem{panichella2013effectively}
A.~Panichella, B.~Dit, R.~Oliveto, M.~Di~Penta, D.~Poshyvanyk, A.~De~Lucia, How
  to effectively use topic models for software engineering tasks? an approach
  based on genetic algorithms, in: Proceedings of the 2013 International
  Conference on Software Engineering, IEEE Press, 2013, pp. 522--531.

\bibitem{cora10}
A.~Corazza, S.~Di~Martino, F.~Ferrucci, C.~Gravino, F.~Sarro, E.~Mendes, How
  effective is tabu search to configure support vector regression for effort
  estimation?, in: Proceedings of the 6th international conference on
  predictive models in software engineering, ACM, 2010, p.~4.

\bibitem{minku2013analysis}
L.~L. Minku, X.~Yao, An analysis of multi-objective evolutionary algorithms for
  training ensemble models based on different performance measures in software
  effort estimation, in: Proceedings of the 9th international conference on
  predictive models in software engineering, ACM, 2013, p.~8.

\bibitem{song2013impact}
L.~Song, L.~L. Minku, X.~Yao, The impact of parameter tuning on software effort
  estimation using learning machines, in: Proceedings of the 9th international
  conference on predictive models in software engineering, ACM, 2013, p.~9.

\bibitem{lessmann2008benchmarking}
S.~Lessmann, B.~Baesens, C.~Mues, S.~Pietsch, Benchmarking classification
  models for software defect prediction: A proposed framework and novel
  findings, IEEE Trans. Softw Eng. 34~(4) (2008) 485--496.

\bibitem{Runeson2009}
P.~Runeson, M.~H\"{o}st,
  \href{http://dx.doi.org/10.1007/s10664-008-9102-8}{Guidelines for conducting
  and reporting case study research in software engineering}, Empirical Softw.
  Engg. 14~(2) (2009) 131--164.
\newblock \href {http://dx.doi.org/10.1007/s10664-008-9102-8}
  {\path{doi:10.1007/s10664-008-9102-8}}.
\newline\urlprefix\url{http://dx.doi.org/10.1007/s10664-008-9102-8}

\bibitem{Cruzes11}
D.~S. Cruzes, T.~Dyba, P.~Runeson, M.~Host,
  \href{http://dx.doi.org/10.1109/ESEM.2011.44}{Case studies synthesis: Brief
  experience and challenges for the future}, in: Proceedings of the 2011
  International Symposium on Empirical Software Engineering and Measurement,
  ESEM '11, IEEE Computer Society, Washington, DC, USA, 2011, pp. 343--346.
\newblock \href {http://dx.doi.org/10.1109/ESEM.2011.44}
  {\path{doi:10.1109/ESEM.2011.44}}.
\newline\urlprefix\url{http://dx.doi.org/10.1109/ESEM.2011.44}

\bibitem{LowryBK98}
M.~Lowry, M.~Boyd, D.~Kulkami, Towards a theory for integration of mathematical
  verification and empirical testing, in: Automated Software Engineering, 1998.
  Proceedings. 13th IEEE International Conference on, IEEE, 1998, pp. 322--331.

\bibitem{rakitin01}
S.~Rakitin, Software Verification and Validation for Practitioners and
  Managers, Second Edition, Artech House, 2001.

\bibitem{krishna2016too}
R.~Krishna, T.~Menzies, W.~Fu, Too much automation? the bellwether effect and
  its implications for transfer learning, in: Proceedings of the 31st IEEE/ACM
  International Conference on Automated Software Engineering, ACM, 2016, pp.
  122--131.

\bibitem{nam2015heterogeneous}
J.~Nam, S.~Kim, Heterogeneous defect prediction, in: Proceedings of the 2015
  10th joint meeting on foundations of software engineering, ACM, 2015, pp.
  508--519.

\bibitem{tan2015online}
M.~Tan, L.~Tan, S.~Dara, C.~Mayeux, Online defect prediction for imbalanced
  data, in: Proceedings of the 37th International Conference on Software
  Engineering-Volume 2, IEEE Press, 2015, pp. 99--108.

\bibitem{rahman14:icse}
F.~Rahman, S.~Khatri, E.~T. Barr, P.~Devanbu, Comparing static bug finders and
  statistical prediction, in: Proceedings of the 36th International Conference
  on Software Engineering, ACM, 2014, pp. 424--434.

\bibitem{scikit-learn}
F.~Pedregosa, G.~Varoquaux, A.~Gramfort, V.~Michel, B.~Thirion, O.~Grisel,
  M.~Blondel, P.~Prettenhofer, R.~Weiss, V.~Dubourg, J.~Vanderplas, A.~Passos,
  D.~Cournapeau, M.~Brucher, M.~Perrot, E.~Duchesnay, Scikit-learn: Machine
  learning in {P}ython, Journal of Machine Learning Research 12 (2011)
  2825--2830.

\bibitem{Witten2011}
I.~Witten, E.~Frank, {Data Mining, practical machine learning tools and
  techniques}, 3rd Edition, Morgan Kaufmann Pub, 2011.

\bibitem{amrit16}
A.~Agrawal, W.~Fu, T.~Menzies, What is wrong with topic modeling? (and how to
  fix it using search-based se), in: arXiv:1608.08176 [cs.SE], 2016, available
  from \url{http://arxiv.org/abs/1608.08176}.

\bibitem{Bergstra2012}
J.~Bergstra, Y.~Bengio, {Random Search for Hyper-Parameter Optimization},
  Journal of Machine Learning Research 13 (2012) 281--305.

\bibitem{kuhn2014caret}
M.~Kuhn, J.~Wing, S.~Weston, A.~Williams, C.~Keefer, A.~Engelhardt, T.~Cooper,
  Z.~Mayer, R.~C. Team, M.~Benesty, et~al., caret: classification and
  regression training. r package version 6.0-24 (2014).

\bibitem{durillo2011jmetal}
J.~J. Durillo, A.~J. Nebro, jmetal: A java framework for multi-objective
  optimization, Advances in Engineering Software 42~(10) (2011) 760--771.

\bibitem{storn1997differential}
R.~Storn, K.~Price, Differential evolution--a simple and efficient heuristic
  for global optimization over continuous spaces, Journal of global
  optimization 11~(4) (1997) 341--359.

\bibitem{scott1974cluster}
A.~Scott, M.~Knott, A cluster analysis method for grouping means in the
  analysis of variance, Biometrics (1974) 507--512.

\bibitem{efron93}
B.~Efron, R.~J. Tibshirani, An introduction to the bootstrap, Mono. Stat. Appl.
  Probab., Chapman and Hall, London, 1993.

\bibitem{Vargha00}
A.~Vargha, H.~D. Delaney, A critique and improvement of the cl common language
  effect size statistics of mcgraw and wong, Journal of Educational and
  Behavioral Statistics 25~(2) (2000) 101--132.

\bibitem{mittas13}
N.~Mittas, L.~Angelis, Ranking and clustering software cost estimation models
  through a multiple comparisons algorithm, IEEE Trans. Software Eng. 39~(4)
  (2013) 537--551.

\bibitem{arcuri11}
A.~Arcuri, L.~Briand, A practical guide for using statistical tests to assess
  randomized algorithms in software engineering, in: Software Engineering
  (ICSE), 2011 33rd International Conference on, IEEE, 2011, pp. 1--10.

\bibitem{papa13}
V.~Papakroni, Data carving: Identifying and removing irrelevancies in the data,
  Master's thesis, Lane Department of Computer Science and Electrical
  Engineering, West Virginia Unviersity (2013).

\bibitem{province15}
B.~Province, Exploiting low dimensionality in software engineering, Master's
  thesis, CSEE, West Virginia University (2015).

\bibitem{aha1991instance}
D.~W. Aha, D.~Kibler, M.~K. Albert, Instance-based learning algorithms, Machine
  learning 6~(1) (1991) 37--66.

\bibitem{grassberger1983measuring}
P.~Grassberger, I.~Procaccia, Measuring the strangeness of strange attractors,
  Physica D: Nonlinear Phenomena 9~(1) (1983) 189--208.

\bibitem{levina04}
E.~Levina, P.~J. Bickel, Maximum likelihood estimation of intrinsic dimension,
  in: Advances in neural information processing systems, 2004, pp. 777--784.

\bibitem{me99q}
T.~Menzies, B.~Cukic, {When to Test Less}, IEEE Software 17~(5) (2000)
  107--112.

\end{thebibliography}

\end{document}